\documentclass[prl,reprint,aps,amsmath,amssymb,twocolumn,superscriptaddress,longbibliography,preprintnumbers]{revtex4-1}
\usepackage{graphicx,bm,color,appendix}
\usepackage[colorlinks,bookmarks=true,citecolor=blue,linkcolor=blue,urlcolor=blue]{hyperref}
\usepackage{times}
\usepackage{amsmath,amssymb}
\usepackage{ytableau} 
\usepackage{booktabs}
\usepackage{comment}
\usepackage{appendix}

\usepackage{xcolor}
\definecolor{darkblue}{rgb}{0.,0.,0.4}
\definecolor{darkred}{rgb}{0.5,0.,0.}
\definecolor{BlueViolet}{RGB}{138,43,226}
\definecolor{SkyBlue}{RGB}{30,144,255}
\definecolor{DarkGreen}{RGB}{0,100,0}

\usepackage[normalem]{ulem}
\usepackage{comment}

\newcommand{\Dh}{\hat\Delta}
\newcommand{\Norb}{N_m}
\newcommand{\Ltop}{\mathrm{U(1)}_{\rm top}}

\begin{document}

\preprint{KYUSHU-HET-383}

\title{
Accessing Continuous Monodromy Defects of the 3D Conformal Field Theory on the Fuzzy Sphere
}

\author{Yin Tang}\email[]{tang.yin.129@m.kyushu-u.ac.jp}
\affiliation{Institute for Advanced Study, Kyushu University, Fukuoka 819-0395, Japan}

\author{W. Zhu}
\email{zhuwei@westlake.edu.cn}

\affiliation{Department of Physics, School of Science, Westlake University, Hangzhou 310030, China }

\begin{abstract}

Monodromy defects are line defects at the boundary of a topological symmetry surface, around which charged operators acquire a symmetry phase. They probe how a critical system responds to an inserted symmetry flux, but their universal information at strongly coupled fixed points, in particular for a continuous family of twists, remains hard to access.
Here we investigate the continuous family of $U(1)$ monodromy defect with fractional twist angles embedded in the three-dimensional $O(2)$ Wilson--Fisher conformal field theory. 
We realize this continuous family of defects within the fuzzy-sphere framework by threading a fractional Aharonov--Bohm flux through the two poles of the sphere. 
We observe charged defect operators carrying fractional transverse spins, as a characteristic signature of the imposed monodromy, and a displacement operator close to its protected dimension, supporting the conformal nature of the defect line.
Crucially, we resolve no relevant symmetry-preserving scalar, and identify the charge-one defect operators as the only relevant, symmetry-breaking deformations.
Assuming the particle--vortex duality, the U(1) monodromy defect maps to a Wilson line of fractional charge in the critical noncompact Abelian Higgs model, and our results then provide evidence that these Wilson lines with fractional charge are locally stable conformal lines.
\end{abstract}

\maketitle

\begin{figure}[t]
\centering
\includegraphics[width=\linewidth]{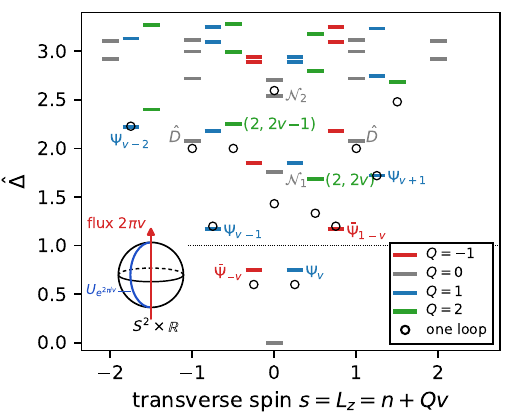}
\caption{
Monodromy defect spectrum of O(2) CFT at $v=\frac14$, $\Norb=13$, versus the transverse spin $s=L_z=n+Qv$ (using bulk-velocity normalization). The identified primaries are labeled. 
(Inset) Schematic plot of realization of the monodromy defect by threading a fractional Aharonov--Bohm flux $2\pi v$ at the two poles of fuzzy sphere (see a companion paper \cite{tang2026monodromydefects3dwilsonfisher} for detail).
}
\label{fig:spec}
\end{figure}

\textit{Introduction.---} 
Line and loop operators extend the local operator content of a quantum field theory and encode the response of critical matter to impurities, boundaries and external probes.
Among them, monodromy (twist) defects are singled out by symmetry: a monodromy defect $\mathcal D_g$ is the boundary of the topological surface operator of a global symmetry element $g$~\cite{Gaiotto:2014kfa}, so that operators charged under $g$ pick up the phase $g$ when transported around the line. For a U(1) symmetry the twist $e^{2\pi iv}$ is continuous; the defect operators then carry fractional transverse spin $s\in\mathbb Z+v$~\cite{Wilczek:1982wy,Arovas:1984qr}, and the defect describes the insertion of a fractional symmetry flux (see the inset of Fig.~\ref{fig:spec}), such as a vortex line in a superfluid or an Aharonov--Bohm flux threading critical matter~\cite{Aharonov:1959fk,Alford:1988sj}. 
In particular, under particle--vortex duality, the U(1) monodromy defects of the O(2) conformal field theory (CFT) become Wilson lines of fractional charge in the critical Abelian Higgs model~\cite{Metlitski:2007fq}.

The last correspondence provides one of the main motivations for the present work. Wilson lines are gauge-invariant extended observables of quantum matter with emergent gauge fields~\cite{Wilson:1974sk,Fradkin:1978dv,Savary:2016ksw}: in gapped topological phases they represent anyon worldlines~\cite{Witten:1988hf,Kitaev:1997wr,Nayak:2007ln}, while at a critical point the renormalized line probes fractionalization, deconfinement and the screening of an external charge~\cite{POLYAKOV1978477,SVETITSKY1982423,Pisarski2006,Senthil:2004dqcp,Senthil:2004lgw,Wang:2017txt,brennan2025,Aharony:2022ntz,Aharony:2023amq,Gabai:2022vri,Gabai:2022mya}. Their universal information in strongly coupled gauge theories is hard to obtain, since perturbative treatments are limited and numerical calculations must resolve the response to a fluctuating gauge field~\cite{Kajantie:2004vy,Motrunich:2004hh,Kragset:2006ept,Bonati:2020jlm}. For the noncompact Abelian Higgs (AH) model with one unit-charged scalar, whose critical point is mapped by particle--vortex duality to the O(2) Wilson--Fisher CFT~\cite{Peskin:1977kp,Dasgupta:1981zz,Fisher:1989bos,Kajantie:2004vy,Karch:2016sxi,Seiberg:2016gmd}, a Wilson line of charge $q$ maps to the monodromy defect $\mathcal D_v$ with $v=q\bmod1$~\cite{Metlitski:2007fq}. Therefore, knowledge of $\mathcal D_v$, obtained on the scalar side without reference to the gauge field, translates into statements about Wilson lines to the extent that the duality holds.
However, even on the scalar side, monodromy defects with continuous $v$ are known only from the $\epsilon$-expansion~\cite{Soderberg:2017oaa,Giombi:2021uae,MCheng2026a} and from large-$N$ and free-field analyses~\cite{Giombi:2021uae,Bianchi:2021snj,Hung:2014npa,Bianchi:2015liz}; non-perturbative conformal data exist for discrete twists~\cite{Billo:2013jda,Gaiotto:2013nva,Gimenez-Grau:2021wiv,Gimenez-Grau:2022czc,ambrosino2026monodromydefectschernsimonstheory} but remain limited for continuous $v$.

In this paper, we study the U(1) monodromy defects of the O(2) CFT using the recently developed fuzzy-sphere framework~\cite{Zhu:2022gjc,He:2026fsj}. 
Fuzzy sphere is a non-perturbative method to access the conformal data including scaling dimensions \cite{Zhu:2022gjc} and operator product expansion (OPE) coefficients ~\cite{Hu:2023xak,Han:2023yyb,Hu:2024pen,Fardelli:2024qla,Fan:2024vcz}, which has been successfully applied to investigate various strongly-correlated 3D CFTs including
O($N$) Wilson--Fisher CFTs~\cite{Han:2023lky,Guo:2025onf,Dey:2025o3,Dey:2026o2,Stergiou:2026cub}, Lee-Yang non-unitary CFT \cite{ArguelloCruz:2025zuq,Fan:2025bhc,EliasMiro:2025msj}, deconfined
criticality~\cite{Zhou:2023qfi,Zhou:2024spn,Yang:2026cof,Chen:2023xjc}, free scalar CFT \cite{He:2025ong,Taylor:2025odf}, topological phase transitions~\cite{Voinea:2026maj,Zhou:2025fer,Gao:2025gny,Zhou:2025csm}, 
superconformal field theories \cite{Zhou:2025fer,Tang:2026sci}, surface CFTs~\cite{Zhou:2024dbt,Dedushenko:2024nwi,Feng:2026sur}, defect CFTs~\cite{Hu:2023ghk,Zhou:2023fqu,Cuomo:2024psk,Yang:2026onl,Liu:2026dqc,Sarma:2026bk}. 
In a companion paper~\cite{tang2026monodromydefects3dwilsonfisher}, we integrated $\mathbb Z_2$ twist defects into the fuzzy-sphere framework by threading an Aharonov--Bohm $\pi$-flux~\cite{Aharonov:1959fk} through the poles of the sphere, and here we extend this route to continuous twists.

Specifically, we realize the U(1) monodromy defect $\mathcal D_v$ of 3D O(2) CFT, with the microscopic couplings kept at the bulk critical point, and obtain the conformal data (e.g. scaling dimensions and operator product expansion (OPE) coefficients) regarding this continuous twist defect. 
We observe defect operators with fractional transverse spins, as a direct signature of the monodromy. 
Through the comparison with perturbative results, we present a detailed analysis of interaction corrections to the scaling dimensions of the identified operators including the displacement operator and charged operators. 
In particular, no relevant perturbation preserving the symmetries of the line is resolved, and the only relevant deformations resolved are the charge-one defect operators. Assuming particle--vortex duality, these results translate into evidence that Wilson lines of the critical AH model are locally stable conformal lines at the sampled charges.

\textit{Monodromy defects and their dual  description}.---
A U(1) monodromy defect $\mathcal D_v$ of the 3D O(2) CFT is the boundary of the U(1) symmetry surface $U_{e^{2\pi iv}}$~\cite{Gaiotto:2014kfa} (see the inset of Fig.~\ref{fig:spec}); it inserts a background $U(1)$ holonomy $2\pi v$ around a straight line $L$, so that a bulk operator $\mathcal O$ of global charge $Q$ picks up the phase:
\begin{align}\label{eq:def}
\mathcal O(r,\varphi+2\pi,z)
=e^{2\pi i v Q} \mathcal O(r,\varphi,z).
\end{align}
Correlators involving charged insertions are defined with a chosen branch surface ending on $L$, and moving the surface across a charged insertion multiplies the correlator by a phase. Integer $v$ gives the trivial defect, while noninteger $v$ enforces a nontrivial twist.

Under particle--vortex duality~\cite{Peskin:1977kp,Dasgupta:1981zz,Fisher:1989bos,Karch:2016sxi}, the O(2) CFT describes the critical point of the $(2+1)$-dimensional noncompact AH model with one complex scalar of unit gauge charge: the topological current $j_{\mathrm{top}}^\mu=(2\pi)^{-1}\epsilon^{\mu\nu\rho}\partial_\nu a_\rho$ is identified with the global $U(1)$ current, and the minimal $2\pi$ monopole operator with the charge-one order parameter $\phi=(\phi_1+i\phi_2)/\sqrt2$~\cite{Borokhov:2002ib,Pufu:2013vpa}. A straight Wilson line $W_q(L)=\exp(iq\int_L a_\mu dx^\mu)$, an infinitely heavy external charge $q$ coupled to the gauge field $a_\mu$, then maps to $\mathcal D_v$ with $v\equiv q\pmod1$ on the O(2) CFT side~\cite{Metlitski:2007fq}. When the bulk is critical, interactions localized on the line can undergo RG flow, describing the polarization and screening of the external charge by the surrounding matter~\cite{Aharony:2022ntz,Aharony:2023amq}. Unit-charged matter can screen integer charges but cannot remove a fractional one~\cite{Metlitski:2007fq}. In what follows we work entirely on the O(2) CFT side and return to the gauge-theory interpretation at the end.

At a conformal defect fixed point, the bulk-to-defect OPE regarding the charge-1 operator takes the form
\begin{align}\label{eq:ope}
\phi(r,\varphi,z)
=\sum_{s}\sum_\alpha
b_{s,\alpha}\,r^{\widehat\Delta_{s,\alpha}-\Delta_\phi}
e^{is\varphi}\hat \Psi_{s,\alpha}(z)+\cdots,
\end{align}
where defect operators $\hat \Psi_{s,\alpha}$ carry fractional transverse spin $s\in\mathbb Z+v $ and have scaling dimension $\hat \Delta_{s,\alpha}$ ($\alpha$ distinguishes  primaries with the same transverse spin)\cite{Giombi:2021uae}. OPE coefficient $b_{s,\alpha}$ measures the coupling of the bulk field to the corresponding defect primary. 
Breaking of translational invariance perpendicular to the line dictates the existence of the displacement operator $\hat D$, i.e. a charge neutral singlet with spin $s = \pm1$ whose scaling dimension is
protected at 2 by the Ward identity~\cite{Billo:2016cpy,Liendo:2012hy,Jensen:2015swa}.
Additionally, local RG stability is controlled by symmetry-preserving perturbations, i.e. a sufficient condition for local RG stability is that every nontrivial defect scalar compatible with all preserved symmetries has $\hat\Delta>1$~\cite{Cuomo:2021rkm}.

\textit{Model and method.---}
We use the fuzzy-sphere regularization \cite{Zhu:2022gjc,He:2026fsj} to simulate the U(1) monodromy defect.
In this framework, interacting
fermions live on a Haldane two-sphere with $\Norb=2s_M+1$ degenerate Landau orbitals ($s_M$ is monopole charge at the center of the sphere)~\cite{Haldane:1983xm}.
Accordingly, a local fermion field is expanded as 
$\psi_\alpha(\mathbf{x})=R^{-1}\sum_m Y^{(s_M)}_m(\hat{\mathbf n})c_{m,\alpha}$, where $c_{m,\alpha}$ annihilates a fermion on a Landau orbital $m =-s_M,\dots,s_M$, $R=\sqrt{\Norb}$ is the radius of the sphere and $Y^{(s_M)}_m(\theta,\varphi)$ is a monopole harmonics \cite{Wu:1976ge}
\begin{equation}
Y^{(s_M)}_m(\theta,\varphi)\propto e^{im\varphi}\cos^{s_M+m}\tfrac{\theta}{2}\,\sin^{s_M-m}\tfrac{\theta}{2}.
\label{eq:orb}
\end{equation}
To study 3D O(2) bulk criticality, we consider the three-flavour model~\cite{Guo:2025onf}
\begin{equation}
H=\int d^2\mathbf x\,\Big[n_{\rm tot}^2-U_1\sum_{a=1,2}V_a\nabla^2V_a+h_1\,\psi_0^\dagger\psi_0\Big],
\label{eq:H}
\end{equation}
where $\psi_0$ is a singlet flavor,  $\psi_{a=1,2}$ form an O(2) vector, whose combinations $\psi_\pm=(\psi_1\pm i\psi_2)/\sqrt2$ carry U(1) charge $Q=\pm 1$, and the field $V_a$ is defined as $V_a=\psi_0^\dagger\psi_a+\psi_a^\dagger\psi_0$. 
The density interaction suppresses charge fluctuations, while the remaining terms tune the competition between the singlet-polarized state and $O(2)$ symmetry broken phase. At criticality, $V_1+iV_2=\sqrt2\,(\psi_0^\dagger\psi_++\psi_-^\dagger\psi_0)$, which carries U(1) charge $+1$, has a leading overlap with the charge-one CFT field $\phi$, and the identity-subtracted part of $\psi_0^\dagger\psi_0$ with the singlet $S$.
In Eq.~\eqref{eq:H} the orbitals are unit normalized, $\psi_\alpha=R^{-1}\sum_m Y^{(s_M)}_m c_{m,\alpha}$, $h_1$ multiplies $\psi_0^\dagger\psi_0$ directly, and the one-body terms generated by the self-contractions of the quartic terms are retained (Sec.~B of the Supplemental Material). In this convention we fix $(U_1,h_1)=(0.270,-0.084)$~\cite{Guo:2025onf}, for which the bulk spectrum at $N_m=13$ gives $\Delta_\phi=0.513$, $\Delta_S=1.50$ and $\Delta_t=1.27$, close to the bootstrap values~\cite{Chester:2019ifh}, and keep them fixed for all defects.

The monodromy is introduced by imposing a twisted boundary condition on the orbitals~\cite{Laughlin:1981jd,Arovas:1984qr}. With $\psi_\alpha=R^{-1}\sum_m Y^{(s_M)}_m c_{m,\alpha}$ and $Y^{(s_M)}_m\propto e^{im\varphi}$, the condition that, relative to the common monopole gauge phase, the charged fields acquire the additional phase $e^{2\pi ivQ_\alpha}$ upon encircling the axis, where $Q_\alpha$ is the U(1) charge of the field ($\psi_\pm\to e^{\pm i\alpha}\psi_\pm$, $Q_0=0$), shifts the angular-momentum grid of each flavour by $vQ_\alpha$, $m\in-s_M+vQ_\alpha+\mathbb Z$, and for $0<v<1$, regularity at both poles selects: 
\begin{equation}
\begin{aligned}
&\psi_+:\ m\in\{-s_M+v,\,-s_M+v+1,\,\dots,\,s_M-1+v\},\\
&\psi_-:\ m\in\{-s_M+1-v,\,-s_M+2-v,\,\dots,\,s_M-v\},\\
&\psi_0:\ m\in\{-s_M,\,-s_M+1,\,\dots,\,s_M\}.
\end{aligned}
\label{eq:orbs}
\end{equation}
At $v=\frac12$ these grids are shifted by half a unit relative to the untwisted grid and reproduce the orbital sets of the $\mathbb Z_2$ twist defects in Ref.~\cite{tang2026monodromydefects3dwilsonfisher}.  
The common monopole gauge phase cancels in bilinears, giving
$(V_1+iV_2)(\varphi+2\pi)=e^{2\pi iv}(V_1+iV_2)(\varphi)$ and thus satisfying the twisted condition Eq. \eqref{eq:def}.
Modes within each flavour are orthonormal because their azimuthal
quantum numbers differ by integers and their polar overlaps are Beta integrals (Supple. Mat. Sec.A \cite{sm}). Regularity requires $s_M\pm m\geq0$, leaving $2s_M$ modes per charged flavour and $2s_M+1$ singlet modes. 
The regular orbital sets specify the microscopic defect core, the lattice counterpart of the standard boundary condition on the charged modes~\cite{Giombi:2021uae,Bianchi:2021snj}. Our spectra are consistent with a flow to the standard interacting monodromy defect, which is infrared stable within the perturbative RG analysis of fully twisted defects~\cite{Giombi:2021uae,MCheng2026a}: every charged defect operator lies above $\Delta_\phi$.

We diagonalize the Hamiltonian in sectors of the conserved flavour charge $Q_F=N_+-N_-$ and of the $z$-component of the angular momentum $L_z$. With the field convention above, a defect operator of U(1) charge $Q$ and transverse spin $s$ creates from the defect vacuum a state in the sector $(Q_F,L_z)=(-Q,-s)$; since the sectors $(Q_F,L_z)$ and $(-Q_F,-L_z)$ are degenerate (see below), all figures and tables are labeled by the operator quantum numbers $(Q,s)$.
The combined transformation $C\,R_y(\pi)$ preserves the defect background for every $v$ and relates the sectors $(Q_F,L_z)$ and $(-Q_F,-L_z)$, which are therefore degenerate.
Since $s_M N_e=\Norb(\Norb-1)/2$ (at $N_e=\Norb$) is an integer, the allowed
sectors obey $L_z\in\mathbb Z+vQ_F$, matching the fractional transverse
spins of the monodromy defect. Within the computational ability, we can access the systems with $\Norb=5$--$13$,
$Q=0,\pm1,2$, and
$v\in\{1/12,1/6,1/4,1/3,5/12,1/2\}$.
If the line reaches a conformal fixed point, its dimensions follow
from $\widehat\Delta_a=(R/c)(E_a-E_0)$ \cite{Cardy:1984rp,Cardy:1985lth,Hu:2023ghk}, with $E_0$ the defect
ground-state energy and $c$ calibrated independently from the bulk
spectrum. A displacement level at $\hat\Delta_D=2$ then provides a consistency check, which is also another normalization choice. Matrix elements within the defect
Hilbert space yield bulk-to-defect and defect OPE data \cite{Hu:2023xak,Hu:2023ghk}. Finite-size extrapolations and their
sensitivity to the fit range, form and energy normalization
are discussed in Supple. Mat. Sec.C \cite{sm}.

\begin{figure}[t]
\centering
\includegraphics[width=\linewidth]{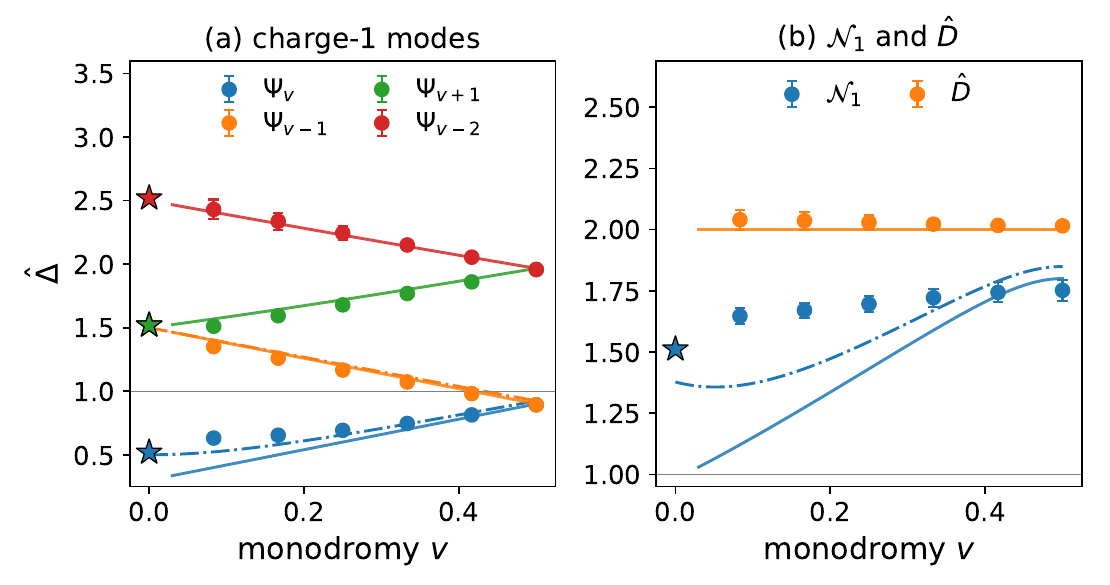}
\caption{Spectral flow of the defect primaries with the monodromy $v$: (a) charge-1 modes $\Psi_s$, $s\in\mathbb Z+v$; (b) lowest neutral scalar $\mathcal N_1$ and displacement $\hat D$. Filled circles: extrapolated data from fuzzy sphere; solid lines: fixed-order one loop [Eq.~\eqref{eq:oneloop}; for $\mathcal N_1$ the one-loop curve is obtained by diagonalizing the truncated mixing matrix, Eq.~(S12)]; dash-dotted: one loop with the resummed defect mass [Eq.~\eqref{eq:resum}]; stars: the corresponding operators of the trivial defect at $v=0$.
The quoted uncertainties reflect the spread of finite-size fits and energy normalizations rather than statistical confidence intervals (see Supple. Mat. Sec. C\cite{sm}).
}
\label{fig:flow}
\end{figure}

\textit{Results.---} Fig.~\ref{fig:spec} displays the typical operator spectrum induced by the monodromy defect with $v=1/4$, organized by the global charge $Q$ and transverse spin $s$. 
A characteristic feature is the appearance of defect operators $\hat \Psi_{s}$ with fractional transverse spins at $s=\frac14,-\frac34,\frac54,-\frac74$, serving as a direct consequence of U(1) monodromy defect (see Eq. \eqref{eq:ope}). The other low-lying operators include the displacement operator $\hat D$ at $s=\pm1$, the neutral scalars $\hat {\mathcal N}_{1,2}$ and charge-two operators labeled by $(2,s)$.  
We can resolve these defect operators and their first-level descendants, supporting the conformal nature of this monodromy defect. 
The aforementioned feature is robust across different twists 
from $v=1/12$ to $1/2$. Fig.~\ref{fig:flow} summarizes the trends of identified primaries, and the extrapolated data are listed in Tab. S1 of the Supple. Mat. \cite{sm}.

The displacement operator provides an independent check of the conformal interpretation. With the energy scale fixed by the bulk velocity alone, the extrapolated dimension of the displacement operator lies between $2.035(4)$ and $2.045(28)$ across the sampled twists [Fig.~\ref{fig:flow}(b)], within $3\%$ of the Ward-identity value $\hat\Delta_{\hat D}=2$. The appearance of the displacement operator close to its protected dimension, without any tuning of defect couplings, together with the descendant structure of the spectrum, supports that the RG flow ends at a conformal defect.

For the charge-one operators, the existing one-loop $\epsilon$-expansion evaluated
at $\epsilon=1$ predicts ~\cite{Wilson:1971dc,Pelissetto:2000ek,Soderberg:2017oaa,Giombi:2021uae}
\begin{equation}
\hat\Delta_{\Psi_s}^{\text{1-loop}}
=\frac12+|s|-\frac{v(1-v)}{5|s|},
\qquad s\in\mathbb Z+v .
\label{eq:oneloop}
\end{equation}
The negative correction lowers the dimensions relative to the
free-scalar value $1/2+|s|$. 
As shown in Fig. \ref{fig:flow}(a), the central estimates of the modes with larger transverse spin follow this expression, although their fit-form uncertainties exceed the quoted spreads: those of
$\hat \Psi_{v+1}$ and $\hat \Psi_{v-2}$ differ from the prediction in Eq. \eqref{eq:oneloop} by less than $0.06$ at all
sampled twists, and those of $\hat \Psi_{v-1}$ differ by less than $0.04$ for
$v\geq1/4$. 
In contrast, the lowest branch $\hat \Psi_{v}$ displays the clearest departure from this approximation, especially for small twists.
This growing discrepancy at small $v$ is consistent with the nonuniformity of the fixed-order expansion: Eq.~\eqref{eq:oneloop} neglects the defect mass coupling $h|\hat\Psi_v|^2$, which is classically irrelevant for $v>0$ but nearly marginal for $v\lesssim\epsilon$, where it must be retained~\cite{MCheng2026a}. We perform a one loop calculation with the resummed defect mass (Supple. Mat. Sec. D \cite{sm}) and obtain a resummed estimate for the lowest charged dimension (the second branch follows by $v\to1-v$)
\begin{equation}
\hat\Delta_{\Psi_v}^{\text{1-loop}'}=1-\tfrac{7\epsilon}{10}+\sqrt{\big(1+\tfrac{2\epsilon}{5}\big)v^2+\tfrac{\epsilon^2}{25}},
\label{eq:resum}
\end{equation}
which, expanded to first order in $\epsilon$ at fixed $v>0$, reproduces Eq.~\eqref{eq:oneloop}, and which approaches the bulk value $\Delta_\phi=1-\frac\epsilon2$ of the trivial defect as $v\to0$. The resummation substantially reduces the discrepancy of the lowest branch for $v\le1/3$ (e.g.\ $0.742$ versus $0.746(21)$ at $v=1/3$, compared with the fixed-order $0.700$).

The charged dimensions also identify relevant defect deformations. The lowest mode
$\hat \Psi_v$ is always relevant with $\hat\Delta_{\Psi_{v}} <1$ throughout all twists. The
second branch has $\hat\Delta_{\Psi_{v-1}}=1.073(13)$ at $v=1/3$ and $0.981(11)$ at $v=5/12$, and linear interpolation places its crossing near $v_c\simeq0.40(2)$, where the error reflects the spread of fit forms and energy normalizations. Within the sampled range $0<v\le\frac12$, the data thus indicate one relevant charge-one branch for $v<v_c$ and two for $v>v_c$, each accompanied by its Hermitian conjugate; the crossing changes the number of relevant symmetry-breaking directions and does not signal a transition of the symmetry-preserving defect.

In the neutral scalar sector, the lowest scalar $\mathcal N_1$ increases
from $\hat\Delta_{\mathcal N_1} =1.696(32)$ at $v=1/4$ to $1.752(42)$ at $v=1/2$, remaining well
above the relevance threshold (Fig. \ref{fig:flow}(b)). The second identified scalar has
dimension $\hat\Delta_{\mathcal N_2}=2.482(44)$ at $v=1/3$ and $2.413(41)$ at $v=1/2$.
At the half-twist, $\mathcal N_1$ is assigned to the $C$-odd channel
and is itself excluded from perturbations preserving $C$. $\mathcal N_1$ and $\mathcal N_2$ become the $C$-odd $\mathcal A$ and $C$-even $\mathcal S$ of the $\pi$-monodromy defect~\cite{tang2026monodromydefects3dwilsonfisher}.
Thus, among the resolved operators, (except for identity) no neutral scalar
compatible with the preserved symmetries is relevant. This supports
stability against symmetry-preserving local perturbations for U(1) monodromy
twists at the sampled $v$. 
In addition, the lowest charge-two operator is also irrelevant (Sec. E of Supple. Mat. \cite{sm}).
Thus the relevant directions belong to the charge-one sector, despite the presence of relatively low-lying composite excitations.

\begin{figure}[t]
\centering
\includegraphics[width=\linewidth]{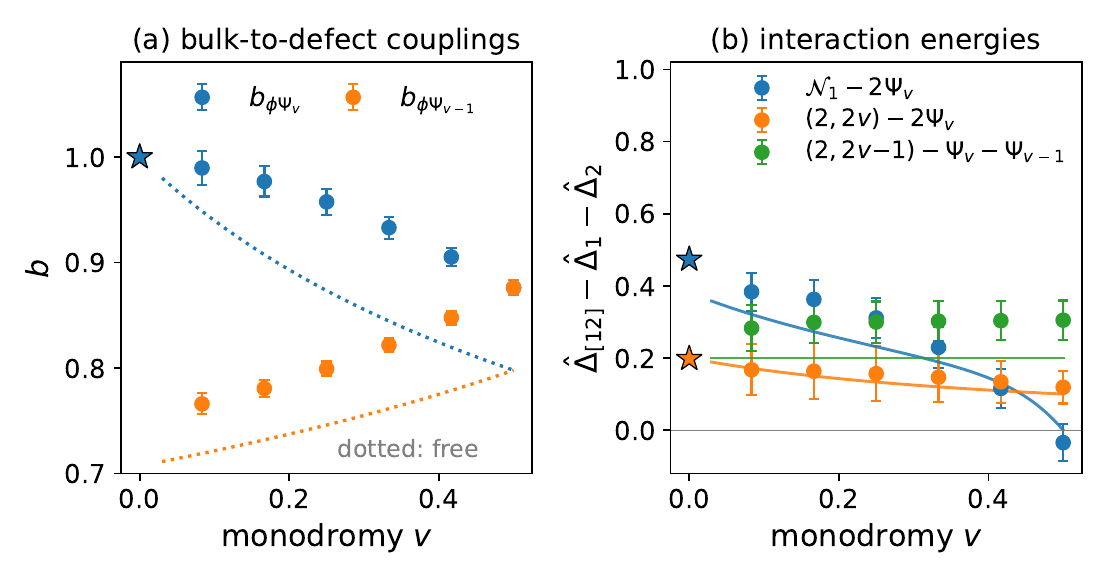}
\caption{(a) Bulk-to-defect couplings of $\phi$ to the two lowest charge-1 modes from fuzzy sphere (dotted: free theory~\cite{Gaiotto:2013nva,Giombi:2021uae,Bianchi:2021snj}; star: trivial defect). (b) Interaction energies $\Dh_{[\hat O_1\hat O_2]}-\Dh_{\hat O_1}-\Dh_{\hat O_2}$ of two-mode operators. Solid lines: the one-loop contact terms (Sec.D \cite{sm}); Stars: $\Delta_S-2\Delta_\phi$ and $\Delta_t-2\Delta_\phi$ of the trivial defect. 
}
\label{fig:data}
\end{figure}

Moreover, the interaction between defect modes can be further quantified by the two-mode interaction energies: $\Dh_{[\hat O_1\hat O_2]}-\Dh_{\hat O_1}-\Dh_{\hat O_2}$, as shown in Fig.~\ref{fig:data}(b).  In the charge-two spectrum,  the central values of $\Delta_{[2,2v]}-2\Delta_{\Psi_v}$ decrease from $0.158$ to $0.119$ for the twist from $v=1/4$ to $1/2$, largely following the trend from the
one-loop contact contribution. The mixed pair instead gives $\Delta_{(2,2v-1)}-\Delta_{\Psi_v}-\Delta_{\Psi_{v-1}}\simeq0.28$--$0.31$, approximately $50\%$ above the one-loop value  $0.2$. 
These positive shifts correspond to positive interaction shifts in the charge-two composite dimensions. 
For the neutral mode, the two-mode interaction energy
$\hat\Delta_{\mathcal N_1}-2\hat\Delta_{\Psi_v}$ decreases
from $0.312$ at $v=1/4$ to $-0.034$ at $v=1/2$, i.e.\ it nearly cancels at the half-twist. In the neutral sector this difference also contains the mixing between the pairs $\Psi_v\bar\Psi_{-v}$ and $\Psi_{v-1}\bar\Psi_{1-v}$, and should not be read as the contact shift of an isolated pair. The near cancellation at $v=\frac12$ reflects that $\mathcal N_1$ becomes the $C$-odd, flavour-antisymmetric combination, on which the flavour-symmetric quartic vertex has no contact term (Supple.\ Mat.\ Sec.~D \cite{sm}).  
These channel-dependent discrepancies are
important when assessing how accurately contact interactions describe the full defect spectrum.

Additionally, the bulk-to-defect OPE coefficients show a systematic departure from the
free-scalar reference (see Fig.~\ref{fig:data}(a)),
$\bigl(b_s^{\mathrm{free}}\bigr)^2
=\frac{\Gamma(\frac12+|s|)}
{\Gamma(\frac12)\Gamma(1+|s|)}$~\cite{Gaiotto:2013nva,Giombi:2021uae,Bianchi:2021snj}.
We observe an enhancement of the measured amplitudes $b_{\phi\Psi_v}$ and
$b_{\phi\Psi_{v-1}}$ compared to free values by
approximately $4$--$11\%$, although they evolve differently depending on $v$: the first decreases from $0.957(12)$ at $v=1/4$ to $0.876(7)$ at $v=1/2$, while the second increases from $0.799(7)$ to the same half-twist value. 
More explicit data are in Tab. S2 (Supple. Mat. \cite{sm}).
These amplitudes quantify the
contribution of the two lowest charged channels to the expansion of the
bulk order parameter near the line.

\textit{Implications for the dual gauge theory.---} If particle--vortex duality holds, the data above describe the Wilson line $W_q$ with charge $q=v$ in the critical AH model: (i) The lines $W_q$ are conformal at all sampled fractional charges, and we resolve no relevant perturbation preserving 
U(1)$_{\rm top}$ and spatial symmetries. The resolved neutral branch, $\Dh_{\mathcal N_1}\approx1.65$--$1.75$, stays well above marginality at all sampled twists, although the present data do not exclude additional branches or transitions between sampled points. (ii) The relevant operators on the line carry $\Ltop$ charge and fractional transverse spin: they are the monopole operators $\Psi_v$
(relevant for all $v$) and $\Psi_{v-1}$ (relevant for $v\gtrsim v_c\simeq0.40$). Perturbing by a single relevant charge-one branch preserves a diagonal combination of transverse rotations and $\Ltop$ and provides a candidate flow towards a monodromy pinning defect~\cite{Kravchuk:2025mpd,Sachdev:1999imp,Allais:2014fqa,ParisenToldin:2016szc,Cuomo:2021kfm,GimenezGrau:2022mag,Bianchi:2022sbz}; perturbing by both branches simultaneously generally breaks this continuous diagonal symmetry. In the gauge theory, these are Wilson lines dressed by monopole operators.
We find no evidence, in the resolved spectrum, of a screening transition within the fractional window, in contrast with weakly coupled and large-$N_f$ gauge theories, where a defect
operator becomes marginal at a critical charge and larger charges are screened~\cite{Aharony:2022ntz,Aharony:2023amq}: for $N_f=1$ the dual monodromy depends on $q$ only modulo one, although the infrared defect realized for a given holonomy can in addition depend on the defect-core interactions and their RG flow. Several low-lying dimensions remain close to additive estimates, while the interaction corrections show appreciable channel dependence.

\textit{Summary and discussion.---}
We have realized the continuous family of U(1) monodromy defects of the O(2) Wilson--Fisher CFT by threading a fractional Aharonov--Bohm flux through the poles of the fuzzy sphere, and obtained their scaling dimensions, bulk-to-defect couplings and defect OPE coefficients as functions of the twist. Defect operators with fractional transverse spins appear as a direct consequence of the monodromy, and a displacement operator close to its protected dimension supports the interpretation of the infrared limit as a conformal defect. All resolved nontrivial defect scalars compatible with the preserved symmetries lie above the relevance threshold, and the only relevant deformations resolved are the charge-one defect operators. These are statements about the O(2) CFT alone. Assuming particle--vortex duality, they translate into the local stability of Wilson lines of the critical Abelian Higgs model at the sampled charges against perturbations preserving $U(1)_{\mathrm{top}}$ and the spatial symmetries.

The current setup is general: any U(1) subgroup of the flavour symmetry in a fuzzy sphere model can be twisted by introducing a
fractional Aharonov--Bohm flux, which gives access to the full family of U(1) monodromy defects of the O($N$) CFTs \cite{Han:2023lky,Guo:2025onf,Dey:2025o3,Dey:2026o2}, and to monodromy defects of fermionic CFTs \cite{Zhou:2025fer,Tang:2026sci}. For partial monodromies such as $e^{2\pi ivS_z}$ in O(3), the untwisted components admit a classically marginal defect mass that can be generated under RG; pole-localized couplings can modify the defect RG trajectory and reduce crossover effects, and whether tuning is needed depends on the microscopic realization and its basin of attraction~\cite{MCheng2026a}.
Moreover, these non-perturbative data provide concrete targets for the bootstrap of monodromy defects~\cite{Gimenez-Grau:2022czc}.
A future determination of these fractional Wilson line data on the gauge-theory side would provide a further test of the duality for extended operators.

\textit{Note added.---} During the preparation of this work, we became aware of an independent study~\cite{Qian2026} using a similar strategy. Whereas Ref.~\cite{Qian2026} retunes the bulk couplings at each $v$ by optimizing the energy scaling, we keep them at the bulk critical point, so that the infrared defect is selected by the defect RG flow alone. 
Partial monodromy defects of the O($N$) CFTs have also been studied by field theory and numerics in Refs.~\cite{MCheng2026a,Joydeep,MCheng2026b}.

\textit{Acknowledgment.---}	
We acknowledge stimulating discussion with M. Cheng. W.Z. thanks Y. C. He for critical comments. 
This work was supported by NSFC No. 12625407, 12474144. We used ChatGPT (OpenAI, GPT-6 astra) and Claude (Anthropic, Opus 5.5) to assist with drafting, writing the numerical codes and checking the manuscript under our direction. The authors are responsible for all content and conclusions.

\bibliography{u1_monodromy}

\clearpage

\appendix

\makeatletter
\@removefromreset{equation}{section}
\makeatother

\setcounter{equation}{0}
\renewcommand{\theequation}{S\arabic{equation}}

\setcounter{secnumdepth}{1}

\begin{widetext}

\begin{center}
\textbf{Supplementary materials for ``Accessing Continuous Monodromy Defects of the 3D Conformal Field Theory on the Fuzzy Sphere''}    
\end{center}

\setcounter{subsection}{0}
\setcounter{equation}{0}
\setcounter{figure}{0}
\renewcommand{\theequation}{S\arabic{equation}}
\renewcommand{\thefigure}{S\arabic{figure}}
\renewcommand{\thetable}{S\arabic{table}}
\setcounter{table}{0}

\tableofcontents

\vspace{2\baselineskip} 

This Supplemental Material provides details supporting the main text. 
Sec.~\ref{app:orb} discusses the fractional-$m$ orbitals, Sec.~\ref{app:me} the interaction matrix elements at fractional $m$, and Sec.~\ref{app:obs} the observables and the finite-size extrapolations. Sec.~\ref{app:oneloop} presents the one-loop analysis, starting from the perturbative Hamiltonian on $\mathbb R\times S^3$. Sec.~\ref{app:cmp} compares the two energy normalizations and collects the data tables. Sec.~\ref{app:smallv} discusses the $v\to0^+$ crossover.

\section{Fractional-$m$ orbitals}
\label{app:orb}
The lowest Landau level (LLL) of a sphere with monopole charge $s$ is spanned by the zero modes of the covariant lowering operator, which in a gauge regular
away from the poles are $u^{s+m}v^{s-m}$ with $u=\cos\frac\theta2e^{i\varphi/2}$, $v=\sin\frac\theta2e^{-i\varphi/2}$, i.e.\ Eq.~3 in the main text. 
A flat U(1) connection with holonomy $2\pi vQ_\alpha$ around the axis adds the AB vector potential $vQ_\alpha\,d\varphi$ to the covariant derivative
acting on flavour $\alpha$; the zero modes are the same functions with $m\to m+vQ_\alpha$, which are multivalued on the sphere exactly as the
monodromy requires. For a product of two such functions of the same flavour the phases cancel, so all flavour-diagonal densities are
single valued; the bilinears $\psi^\dagger_\pm\psi_0$ carry the winding $e^{\mp 2\pi iv}$.

\emph{Orthonormality.} 
For $m,m'$ in the same grid $-s+vQ_\alpha+\mathbb Z$,
\begin{equation}
\int d\Omega\,\bar\Phi_m\Phi_{m'}=2\pi\delta_{mm'}\mathcal N_m^2\!\int_{-1}^{1}\!d\cos\theta\,u^{s+m}v^{s-m},
\end{equation}
with $u=\cos^2\frac\theta2$, $v=\sin^2\frac\theta2$,
since $m-m'\in\mathbb Z$; the remaining integral is $2B(s+m+1,s-m+1)$, which is finite for $s\pm m>-1$ and is cancelled by $\mathcal N_m^2$.
The sets of LLLs are thus orthonormal. Their wavefunctions vanish at both poles, with the lowest
exponents $1-v$ (north) and $v$ (south) for $\psi_+$ and the reverse for $\psi_-$: the regular behaviour of a charged field at an AB flux.
The two neighbouring members of the $\psi_+$ grid, $m=s+v$ and $m=-s-1+v$, behave as $r^{-v}$ at the north pole and $r^{-(1-v)}$ at the south
pole, respectively; they are square integrable for $0<v<1$ but singular. Admitting one of them corresponds to a different self-adjoint extension of the single-particle problem and changes the microscopic boundary condition; whether the resulting interacting theory flows to the alternative (``$\Delta^-$'') monodromy defect of Refs.~\cite{Giombi:2021uae,Bianchi:2021snj,Breitenlohner:1982jf,Klebanov:1999tb}, which is infrared unstable in the perturbative analysis~\cite{MCheng2026a}, would require a separate spectral analysis. The LLL condition is a first-order equation whose solutions on each grid are the functions $u^{s+m}v^{s-m}$ with $m\in-s+vQ_\alpha+\mathbb Z$; regularity at both poles requires $-s\le m\le s$, i.e.\ $m=-s+v+k$ with $k=0,\dots,2s-1$ for $\psi_+$ (and correspondingly for $\psi_-$), so the regular LLL of a charged flavour has exactly $2s$ states and that of $\psi_0$ has $2s+1$.

\section{Matrix elements at fractional $m$}
\label{app:me}
Write $\Phi_m=\mathcal N_m e^{im\varphi}u^{\alpha_m}v^{\beta_m}$ with $u=\cos^2\frac\theta2$, $v=\sin^2\frac\theta2$, $\alpha_m=\frac{s+m}2$,
$\beta_m=\frac{s-m}2$ (here $u,v$ are real; $v$ should not be confused with the monodromy). In the LLL the interaction Eq. (4) in the main text becomes
$\sum V_{1234}\,c^\dagger_{1}c^\dagger_{2}c_{3}c_{4}$ with two types of integral. The $\delta$-function term needs
\begin{align}
I_\delta&=\int d\Omega\,\bar\Phi_1\bar\Phi_2\Phi_3\Phi_4
=2\pi\,\delta_{m_1+m_2,m_3+m_4}\Big(\prod_i\mathcal N_i\Big)\,2B(P+1,Q+1),
\label{eq:Idelta}
\end{align}
with $P=\sum_i\alpha_i$, $Q=\sum_i\beta_i$, where the $\varphi$ integral imposes conservation of the total $m$, including its fractional
part, and the $\theta$ integral is a Beta function. The $\nabla^2$ term needs $\int d\Omega\,\bar\Phi_1\Phi_4\nabla^2(\bar\Phi_2\Phi_3)$. The product
$\bar\Phi_2\Phi_3=\mathcal N_2\mathcal N_3e^{ik\varphi}u^Av^B$ with $A=\alpha_2+\alpha_3$, $B=\beta_2+\beta_3$ and $k=m_3-m_2$, which is fractional for
$Q$-changing transitions, and for any real $A,B,k$ the Laplacian on the unit sphere satisfies
\begin{align}
e^{-ik\varphi}\nabla^2\big(u^Av^Be^{ik\varphi}\big)=&\;A^2u^{A-1}v^{B+1}+B^2u^{A+1}v^{B-1}-(2AB+A+B)\,u^Av^B-\tfrac{k^2}4\,u^{A-1}v^{B-1},
\label{eq:lap}
\end{align}
as one verifies from $\nabla^2=\frac1{\sin\theta}\partial_\theta\sin\theta\,\partial_\theta+\frac1{\sin^2\theta}\partial_\varphi^2$ with $\partial_\theta u=-\sqrt{uv}=-\partial_\theta v$.
Each of the four terms is again of the form~\eqref{eq:Idelta} with shifted exponents, so
\begin{align}
I_{\nabla}&=2\pi\,\delta_{m_1+m_2,m_3+m_4}\Big(\prod_i\mathcal N_i\Big)\,J,\nonumber\\
J&=2A^2B(P,Q+2)+2B^2B(P+2,Q)\nonumber\\
&\quad-2(2AB+A+B)\,B(P+1,Q+1)-\tfrac{k^2}2\,B(P,Q).
\label{eq:Ilap}
\end{align}
The Beta function in Eq.~\eqref{eq:Idelta} is always finite. In Eq.~\eqref{eq:Ilap}, $P=\alpha_1+\alpha_4+A$ and $Q=\beta_1+\beta_4+B$ are strictly
positive as soon as one of the four orbitals belongs to a charged flavour, since regular charged orbitals have $\alpha,\beta\ge\frac12\min(v,1-v)$;
this is always the case for the $V_a\nabla^2V_a$ term, whose bilinears pair $\psi_0$ with $\psi_\pm$. The terms that would be singular at
$v=0$ have vanishing prefactors ($A^2$, $B^2$ or $k^2$) there, so Eqs.~\eqref{eq:Idelta}--\eqref{eq:Ilap} reduce continuously to the
pseudopotential form of the undefected sphere. The selection rule is implemented by indexing all orbitals on the common grid
$\frac1D\mathbb Z$ with $D={\rm lcm}(2,q)$ for $v=p/q$. We verified Eq.~\eqref{eq:lap} symbolically for real exponents and
Eqs.~\eqref{eq:Idelta}--\eqref{eq:Ilap} against direct numerical quadrature at $v=\frac13$ for $250$ random quadruples of fractional orbitals,
with agreement to $10^{-11}$; the $v=\frac12$ case reproduces term by term the Hamiltonian of the $\pi$-monodromy defect of Ref.~\cite{tang2026monodromydefects3dwilsonfisher}. The alternative, an expansion of the interaction in (twisted) spherical harmonics with ${}_3F_2$ overlaps~\cite{Qian2026}, does not terminate for fractional $m$ because the twisted bilinears are not polynomials on the sphere, and has to be truncated; Eqs.~\eqref{eq:Idelta}--\eqref{eq:Ilap} resum it in closed form. The
one-body terms generated by the self-contractions of Eq.~\eqref{eq:H}, which on the undefected sphere are uniform and amount to a
shift of $h_1$, are evaluated with the defect orbital sets; on the defect they are no longer uniform and act as a perturbation localized
at the poles, consistently with the rest of the Hamiltonian.

\section{Observables and extrapolations}
\label{app:obs}
The defect spectrum is read off from the sectors $(Q,L_z)$ of the twisted Hamiltonian; the bulk spectrum of the same model at the same
$\Norb$ fixes the velocity $c$ from the levels of the conserved current $J$ and the stress tensor $T$~\cite{Cardy:1985lth,Zhu:2022gjc}. Bulk-to-defect couplings are obtained from
$\langle\hat O|\rho(\theta)|0\rangle$, the matrix element of a microscopic density between the defect ground state and a defect eigenstate,
divided by the bulk matrix element of the same density that defines its normalization. Defect conformal symmetry fixes its angular
dependence to $(\sin\theta)^{\Dh-\Delta_O}$, where $\Delta_O$ is the dimension of the leading bulk operator in the density~\cite{Billo:2016cpy};
on the trivial defect ($v=0$) the data reproduce this angular dependence within numerical precision, both for $\phi$ itself and for its transverse derivative
($\Dh=\Delta_\phi+1$). At $v>0$ the residual $\theta$-dependence after this factor is $2$--$3\%$ at $\theta=75^\circ$ and $4$--$8\%$ at
$\theta=60^\circ$ for $\Psi_v$ and $\Psi_{v-1}$ at $\Norb=13$, and it decreases monotonically with $\Norb$. It reflects finite-size corrections
near the poles, and we therefore evaluate $b$ at the equator, the point farthest from the defect. Defect OPE coefficients are obtained from a charged density evaluated near the pole, where it acts as a defect-local operator~\cite{tang2026monodromydefects3dwilsonfisher}: with $|\hat1\rangle,|\hat3\rangle$ defect states and $\hat O_2$ the lowest defect operator transferred by $n_{\mathcal O}$,
\begin{equation}
C_{\hat1\hat2\hat3}=\lim_{\theta_0\to0}\frac{\langle\hat1|n_{\mathcal O}(\theta_0)|\hat3\rangle}{\langle\hat2|n_{\mathcal O}(\theta_0)|\hat0\rangle}\,(\cos\theta_0)^{-(\Dh_1-\Dh_2-\Dh_3)},
\end{equation}
since $\langle\hat O_1^\dagger(\infty)\hat O_2(\tau)\hat O_3(0)\rangle=C_{\hat1\hat2\hat3}\,\tau^{\Dh_1-\Dh_2-\Dh_3}$ with $\tau=\cos\theta_0$, while the normalizing two-point function carries no $\tau$-dependence. The charge and transverse-spin channel of the density is fixed by the selection rules of the states $|\hat1\rangle$, $|\hat3\rangle$, which carry definite $Q_F$ and $L_z$. Corrections of relative order $\theta_0^{\Dh_2'-\Dh_2}$ from the next defect operator in the channel and $\theta_0^2$ from descendants limit the usable window; we average over $\theta_0\in\{15^\circ,30^\circ,45^\circ\}$, where the corrected ratio forms a plateau, and include half of its spread in the error. All OPE coefficients are reported as absolute values here. The one-loop values in Table~\ref{tab:dims} are derived in Sec.~\ref{app:oneloop}; Table~\ref{tab:ope} lists free-theory and GFF reference values.

\emph{Extrapolations.} The extrapolations use the four largest sizes~\cite{Lao:2023zis,Laeuchli:2025fuz}, $\Norb=10$--$13$; the finite-size sequences for $\Norb=5$--$13$ are shown
in Fig.~\ref{fig:fss}. The sequences are smooth, without even--odd effects, and including $\Norb=13$ changes the extrapolated values by
less than $0.003$ for $\Psi_v$, $\Psi_{v-1}$, $\mathcal N_1$, $(2,2v)$ and $\hat D$, and by at most $0.024$ for the higher operators. As a test of the
asymptotic form we also fitted the three-parameter forms $X_\infty+c/\sqrt{\Norb}+c'/\Norb$ and $X_\infty+c/\Norb+c'/\Norb^2$ to the six largest
sizes. For $\Psi_v$, $\Psi_{v-1}$, $\mathcal N_1$, $(2,2v)$ and $\hat D$ they agree with the quoted values within two error bars (deviations $\le0.06$).
For $\Psi_{v+1}$, $\Psi_{v-2}$ and $(2,2v-1)$, whose sequences flatten or bend at the largest sizes, the $1/\sqrt{\Norb}$ three-parameter form lies
lower by up to $0.14$, $0.23$ and $0.12$, respectively; for these operators the quoted errors underestimate the systematic uncertainty. Operators that are degenerate at the free level are identified by the channel in which their OPE coefficient is largest, following Ref.~\cite{tang2026monodromydefects3dwilsonfisher}; all identifications are stable over the full range of sizes.
The estimates for $\Psi_v$, $\Psi_{v-1}$ and $\mathcal N_1$ at $v=\frac1{12}$ and $\frac16$ are in addition sensitive to the small-$v$ crossover of Sec.~\ref{app:smallv}.

\begin{figure*}[t]
\centering
\includegraphics[width=0.9\textwidth]{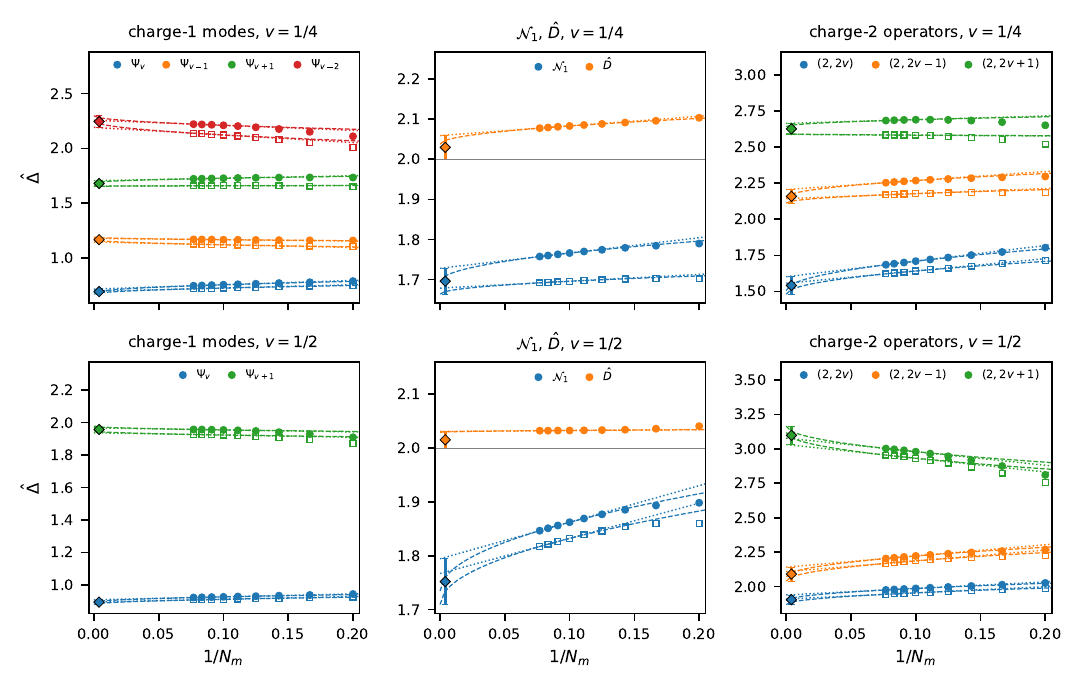}
\caption{Finite-size sequences of the defect dimensions at $v=\frac14$ (top) and $v=\frac12$ (bottom) versus $1/\Norb$ for $\Norb=5$--$13$.
Filled circles: bulk-velocity normalization; open squares: displacement normalization (omitted for $\hat D$, which equals 2 by
construction there); dashed and dotted lines: two-parameter fits in $1/\sqrt{\Norb}$ and $1/\Norb$ to $\Norb=10$--$13$; diamonds: quoted
extrapolations with their errors. At $v=\frac12$, $\Psi_{v-1}$ and $\Psi_{v-2}$ are degenerate with $\Psi_v$ and $\Psi_{v+1}$ and are not shown.}
\label{fig:fss}
\end{figure*}

\section{One-loop analysis}
\label{app:oneloop}
\emph{Perturbative Hamiltonian.} We consider $M=2$ real scalars, $\Phi=(\phi_1+i\phi_2)/\sqrt2$, in $d=4-\epsilon$,
\begin{equation}
S=\int d^dx\Big[\tfrac12(\partial\phi_a)^2+\tfrac{\lambda}{4!}(\phi_a\phi_a)^2\Big],\qquad \frac{\lambda_*}{16\pi^2}=\frac{3\epsilon}{M+8},
\end{equation}
with the monodromy $\Phi(\varphi+2\pi)=e^{2\pi iv}\Phi(\varphi)$, and map $\mathbb R^4$ to the cylinder $\mathbb R\times S^3$, where the defect wraps the great
circle $\chi=0$ of $S^3$ [coordinates $x_1+ix_2=\cos\chi\,e^{i\tau'}$, $x_3+ix_4=\sin\chi\,e^{i\varphi}$]. In the free theory the twisted field is a sum
of modes of transverse spin $s\in\mathbb Z+v$,
\begin{equation}
\Phi=\sum_{s\in\mathbb Z+v}\frac{1}{\sqrt{2\Delta_s}}\Big(a_s\,Y_s+b_s^\dagger\,\bar Y_{-s}\Big)+\dots,\qquad
Y_s=\mathcal N_s(\sin\chi)^{|s|}e^{is\varphi},\quad \mathcal N_s^2=\frac{|s|+1}{2\pi^2},
\end{equation}
where $a_s$ ($b_s$) annihilates a charge $+1$ ($-1$) quantum of energy $\Delta_s=1+|s|$ and the dots denote descendants. The Hamiltonian is
\begin{equation}
H=\sum_{s}\Delta_s\big(a_s^\dagger a_s+b_s^\dagger b_s\big)+\frac{\lambda}{6}\int_{S^3}{:}(\bar\Phi\Phi)^2{:}
+\sum_a\frac{\mu}{2}\int_{S^3}\frac{{:}\phi_a^2{:}}{\sin^2\chi},
\label{eq:Hpert}
\end{equation}
where we used $(\phi\cdot\phi)^2=4(\bar\Phi\Phi)^2$. The last term is generated by normal ordering the quartic vertex with respect to the defect
vacuum: the tadpole $\langle\phi_a\phi_a\rangle_{\mathcal D}=-v(1-v)/(8\pi^2r^2)$ per real component~\cite{Giombi:2021uae,Bianchi:2021snj} gives
\begin{equation}
\mu=\frac{\lambda_*}{6}\Big(\sum_c C_c+2C_a\Big)=-\frac{2}{5}\,\epsilon\,v(1-v),\qquad C_a=-\frac{v(1-v)}{8\pi^2}.
\end{equation}

\emph{Single modes.} The potential term shifts a single mode by $\mu\int|Y_s|^2/\sin^2\chi/(2\Delta_s)=\mu/(2|s|)$, so that
$\Dh_{\Psi_s}=\frac{d-2}{2}+|s|+\frac{\mu}{2|s|}$, which is the one-loop result shown in the main text; equivalently $\Dh_s=\frac{d-2}2+\sqrt{s^2+\mu}$ to this
order~\cite{Soderberg:2017oaa,Giombi:2021uae}.

\emph{Two-mode operators.} The quartic term acts on two-particle states. For four modes with $s_1+s_2=s_3+s_4$ the overlap is
\begin{equation}
\int_{S^3}\frac{Y_{s_1}Y_{s_2}\bar Y_{s_3}\bar Y_{s_4}}{\prod_i\sqrt{2\Delta_{s_i}}}=\frac{1}{4\pi^2\big(2+\sum_i|s_i|\big)},
\end{equation}
and the vertex $\frac{\lambda}{6}(\bar\Phi\Phi)^2$ contracts with a pair of distinct modes in four ways (two for identical modes). At $\lambda_*$ this gives the
contact kernel
\begin{equation}
\gamma^{\rm c}=\frac{3\epsilon}{M+8}\,T\,\frac{2}{1+\frac12\sum_{i=1}^4|s_i|}\times\begin{cases}1&\text{distinct modes},\\ \frac12&\text{identical modes},\end{cases}
\qquad T=\tfrac23 ,
\label{eq:contact}
\end{equation}
which for a diagonal element ($s_3=s_1$, $s_4=s_2$) is $\frac{3\epsilon}{M+8}T\frac{2}{1+|s_1|+|s_2|}$, as for the $\mathbb Z_2$
defects~\cite{tang2026monodromydefects3dwilsonfisher}. At $\epsilon=1$ it gives $0.2/(1+2v)$ for $(2,2v)$, $0.2/(1+v)$ for $(2,2v+1)$, and $0.2$ for
$(2,2v-1)$ and for $\hat D$. Since $\hat D$ and $(2,2v-1)$ consist of the same two modes with the same kernel, both have
$\Dh=\Dh_{\Psi_v}+\Dh_{\Psi_{v-1}}+0.2=2$ for every $v$ at this order. In the neutral $L_z=0$ sector the pairs $\Psi_v\bar\Psi_{-v}$ and
$\Psi_{v-1}\bar\Psi_{1-v}$ mix through the pair-annihilation term ($\sum_i|s_i|=2$), and the lowest two operators are the eigenvectors of
\begin{equation}
\mathsf M(v)=\begin{pmatrix}2\Dh_{\Psi_v}+\dfrac{0.4}{1+2v}&0.2\\[6pt]0.2&2\Dh_{\Psi_{v-1}}+\dfrac{0.4}{3-2v}\end{pmatrix}.
\label{eq:mix}
\end{equation}
Away from $v=\frac12$ the two pairs are not degenerate, and diagonalizing this two-state matrix, whose entries are evaluated to first order in $\epsilon$, retains selected higher-order terms, starting at $O(\epsilon^2/(1-2v))$; the resulting values in Table~\ref{tab:dims} are a truncated-Hamiltonian approximation rather than a strict fixed-order result. At $v=\frac12$ the diagonalization within the degenerate subspace is the standard first-order result.

At $v=\frac12$ the eigenvectors are the $C$-odd combination $\mathcal A\propto\Psi_{1/2}\bar\Psi_{-1/2}-\Psi_{-1/2}\bar\Psi_{1/2}$, with contact term $0$, and
the $C$-even $\mathcal S$, with contact term $0.4$. These are the antisymmetric ($T_A=0$) and singlet ($T_S=\frac43$) flavour channels of the
$\pi$-monodromy defect~\cite{tang2026monodromydefects3dwilsonfisher}: the vertex $(\phi\cdot\phi)^2$ is flavour symmetric and has no contact term in
antisymmetric channels, which is why $\Dh_{\mathcal N_1}-2\Dh_{\Psi_v}$ nearly vanishes at $v=\frac12$.

\emph{Resummed lowest mode.} For $s=v\lesssim\epsilon$ the potential is not small compared with the centrifugal term $v^2$, and the defect mass coupling $h|\hat\Psi_v|^2$, with classical RG eigenvalue $-2v$, becomes nearly marginal and must be kept~\cite{MCheng2026a}. For a fully twisted O(2) defect its one-loop $\beta$-function
[Eq.~(4) of Ref.~\cite{MCheng2026a} at $M=N=2$] is
\begin{equation}
\beta_h=\frac{h^2}{2\pi}+\Big(2v+\frac{2\epsilon}{5}\Big)h+\frac{4\pi\epsilon}{5}\,v(1-v),
\end{equation}
whose discriminant, $4(1+\frac{2\epsilon}5)v^2+\frac{4\epsilon^2}{25}$, is positive for all $v$, so the two fixed points never annihilate. The infrared-stable root, $h_*/2\pi=-v-\frac\epsilon5+\sqrt{(1+\frac{2\epsilon}5)v^2+\frac{\epsilon^2}{25}}=-\frac\epsilon5(1-v)+O(\epsilon^2/v)$, is continuously connected to $h_*=0$ as $\epsilon\to0$ at fixed $v>0$, reproduces the fixed-order anomalous dimension upon expansion, and corresponds to the standard boundary condition. Its use at $\epsilon=1$ and finite $v$ is an extrapolation of the small-$v$ RG analysis. It gives
$\Dh_{\Psi_v}=1-\frac\epsilon2+v+h_*/2\pi$, i.e.\ Eq. (7) in the main text, which at $\epsilon=1$ reads $0.3+\sqrt{1.4v^2+0.04}$. By charge conjugation, $\hat\Delta_{\Psi_{v-1}}(v)=\hat\Delta_{\Psi_{1-v}}(1-v)$, so the same expression evaluated at $1-v$ applies to the second branch; for $1-v\ge\frac12$ it differs from the fixed-order result only by $O(\epsilon^2/(1-v))$, and this part of the curve is an extrapolation of the small-argument RG analysis. The composite values obtained by inserting the resummed constituent dimensions into the contact terms (``1-loop$'$'' in Table~\ref{tab:dims}) are heuristic: the contact kernel is not resummed consistently, and we do not apply the substitution to the protected displacement operator, for which it would give $2.05$ instead of $2$ at $v=\frac12$. At $v\to0$ the substitution gives $2\Delta_\phi+\frac\epsilon5=2-\frac{4\epsilon}5\to1.2$ at $\epsilon=1$ for $(2,2v)$, the bulk one-loop $\Delta_t$. The truncated mixing matrix gives $\mathcal N_1\to1.377$ instead of the bulk one-loop $\Delta_S=1.4$; the difference arises from the higher-order terms retained by the diagonalization and is not a controlled correction. Within the same RG approximation, the dimension of $|\hat\Psi_v|^2$ follows from the slope of $\beta_h$ at the fixed point, $\hat\Delta=d-2+\beta_h'(h_*)=2-\epsilon+2\sqrt{(1+\tfrac{2\epsilon}{5})v^2+\tfrac{\epsilon^2}{25}}$, which does reduce to $\Delta_S=2-\tfrac{3\epsilon}{5}$ at $v=0$. The resummation reorganizes $O(\epsilon^2/v)$ terms; at $v=\frac12$ it
differs from the fixed order by $0.024$ for $\Psi_v$.

\section{Energy normalizations and data tables}
\label{app:cmp}
Figure~\ref{fig:flowfull} shows all identified primaries together with their values at the largest size in both energy normalizations, and the comparison with the
free-theory and one-loop results. At $\Norb=13$ the bulk-velocity and displacement
normalizations differ by $1.6$--$4\%$ for $v\ge\frac14$ and by up to $7\%$ at $v=\frac1{12}$. The extrapolations in the two normalizations are
combined in the quoted values (Sec.~\ref{app:obs}). Figure~\ref{fig:N1} shows the composition of the lowest neutral operator $\mathcal N_1$: its OPE coefficient with $\Psi_v\bar\Psi_{-v}$ grows from $0.672(21)$ at $v=\frac12$ to $0.87(4)$ at
$v=\frac1{12}$, that with $\Psi_{v-1}\bar\Psi_{1-v}$ falls from $0.710(3)$ to $0.24(8)$, and its coupling to the bulk singlet grows from zero,
forbidden by $C$ at $v=\frac12$, to $1.02(8)$. At $v=\frac12$ the two coefficients are related by $R_y(\pi)$ and must coincide in magnitude (their relative sign is negative because $\mathcal N_1$ is $C$-odd); their
difference, $0.04$, measures the systematic error of the ratio method. Tables~\ref{tab:dims} and~\ref{tab:ope} collect the extrapolated data.

\begin{figure*}[h]
\centering
\includegraphics[width=\textwidth]{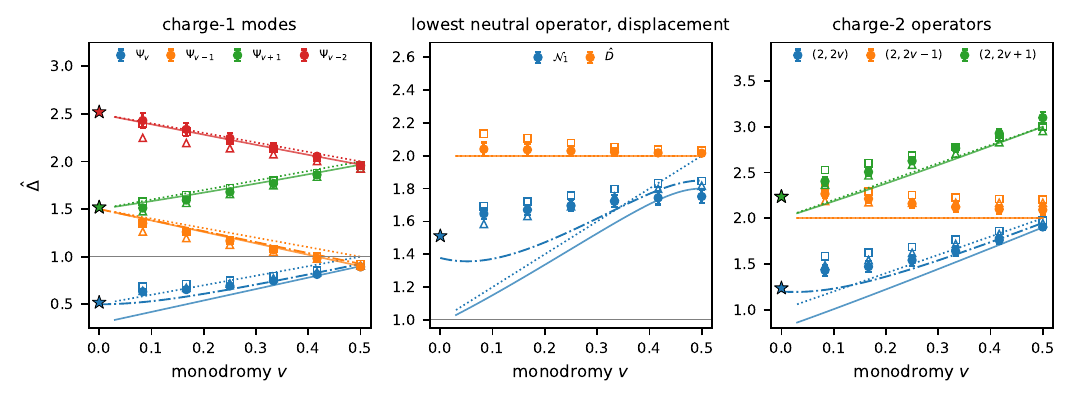}
\caption{Low-lying defect primaries versus $v$. Filled circles: extrapolated data from fuzzy-sphere using $\Norb=10$--$13$; open squares and triangles: raw data from fuzzy-sphere with $\Norb=13$ using the bulk-velocity and displacement normalizations; solid: one loop (Eq.~\eqref{eq:oneloop}; Eq.~\eqref{eq:mix} for $\mathcal N_1$);
dash-dotted: resummed one loop (Eq.~\eqref{eq:resum}); dotted: free theory; stars: results for trivial defect at $v=0$.
}
\label{fig:flowfull}
\end{figure*}

\begin{figure}[h]
\centering
\includegraphics[width=0.45\linewidth]{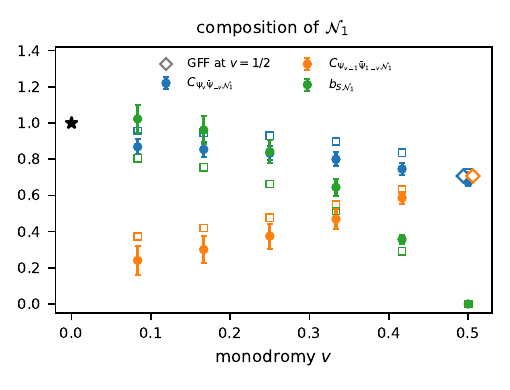}
\caption{Composition of $\mathcal N_1$: its OPE coefficients with the two neutral pairs and its coupling to the bulk singlet $S$ (filled:
extrapolated data from fuzzy-sphere; open: raw data from fuzzy-sphere with $\Norb=13$; diamonds: GFF value $1/\sqrt2$ at $v=\frac12$).}
\label{fig:N1}
\end{figure}


\begin{table*}[h]
\centering\footnotesize
\setlength{\tabcolsep}{5pt}
\begin{tabular}{l|l|c|c|c|c|c|c}
\hline
\hline 
operator & & $v=1/12$ & $v=1/6$ & $v=1/4$ & $v=1/3$ & $v=5/12$ & $v=1/2$\\
\hline
$\Psi_v$ & FS & 0.632(20) & 0.654(23) & 0.692(23) & 0.746(21) & 0.814(18) & 0.893(14)\\
 & 1-loop & 0.400 & 0.500 & 0.600 & 0.700 & 0.800 & 0.900\\
 & 1-loop$'$ & 0.523 & 0.581 & 0.657 & 0.742 & 0.832 & 0.924\\
\hline
$\Psi_{v-1}$ & FS & 1.351(38) & 1.260(31) & 1.167(22) & 1.073(13) & 0.981(11) & 0.893(14)\\
 & 1-loop & 1.400 & 1.300 & 1.200 & 1.100 & 1.000 & 0.900\\
 & 1-loop$'$ & 1.403 & 1.306 & 1.210 & 1.114 & 1.019 & 0.924\\
\hline
$\Psi_{v+1}$ & FS & 1.511(30) & 1.594(29) & 1.679(26) & 1.769(20) & 1.861(16) & 1.958(20)\\
 & 1-loop & 1.569 & 1.643 & 1.720 & 1.800 & 1.882 & 1.967\\
\hline
$\Psi_{v-2}$ & FS & 2.430(78) & 2.337(67) & 2.245(53) & 2.151(39) & 2.055(28) & 1.958(20)\\
 & 1-loop & 2.409 & 2.318 & 2.229 & 2.140 & 2.053 & 1.967\\
\hline
$\mathcal N_1$ & FS & 1.647(33) & 1.670(30) & 1.696(32) & 1.722(37) & 1.744(41) & 1.752(42)\\
 & 1-loop & 1.121 & 1.273 & 1.431 & 1.589 & 1.730 & 1.800\\
 & 1-loop$'$ & 1.364 & 1.432 & 1.542 & 1.670 & 1.790 & 1.849\\
\hline
$\mathcal N_2$ & FS & -- & -- & -- & 2.482(44) & 2.433(42) & 2.413(41)\\
 & 1-loop & -- & -- & -- & 2.423 & 2.273 & 2.200\\
\hline
$(2,2v)$ & FS & 1.431(58) & 1.471(62) & 1.542(60) & 1.639(54) & 1.762(45) & 1.905(35)\\
 & 1-loop & 0.971 & 1.150 & 1.333 & 1.520 & 1.709 & 1.900\\
 & 1-loop$'$ & 1.217 & 1.312 & 1.447 & 1.604 & 1.773 & 1.949\\
\hline
$(2,2v{-}1)$ & FS & 2.266(45) & 2.213(40) & 2.159(47) & 2.121(50) & 2.099(50) & 2.091(51)\\
 & 1-loop & 2.000 & 2.000 & 2.000 & 2.000 & 2.000 & 2.000\\
\hline
$(2,2v{+}1)$ & FS & 2.403(51) & 2.506(48) & 2.627(38) & 2.764(36) & 2.926(49) & 3.098(67)\\
 & 1-loop & 2.154 & 2.314 & 2.480 & 2.650 & 2.824 & 3.000\\
\hline
$\hat D$ & FS & 2.040(41) & 2.045(28) & 2.045(15) & 2.040(5) & 2.035(4) & 2.035(5)\\
 & exact & 2 & 2 & 2 & 2 & 2 & 2\\
\hline\hline 
\end{tabular}
\caption{Defect scaling dimensions of the U(1) monodromy defect $\mathcal D_v$ of the O(2) CFT (under particle--vortex duality, of the Wilson lines $W_v$ of the critical Abelian Higgs model). FS: fuzzy-sphere values extrapolated from $\Norb=10$--$13$ (error bar is from half-spread of the $1/\sqrt{\Norb}$ and $1/\Norb$ fits in
both normalizations);  1-loop: fixed order at $\epsilon=1$ [Eq.~\eqref{eq:oneloop} and Sec.~\ref{app:oneloop}], for $\mathcal N_{1,2}$ the 1-loop values follow from diagonalizing the truncated mixing matrix, Eq.~\eqref{eq:mix}; 1-loop$'$: with the resummed defect mass of the lowest modes [Eq.~\eqref{eq:resum} and Sec.~\ref{app:oneloop}]. 
$\mathcal N_2$ is listed only where it is the $\Psi_{v-1}\bar\Psi_{1-v}$ composite. At $v=\frac12$, $\Psi_{v-1}$ and $\Psi_{v-2}$ are degenerate with
$\Psi_v$ and $\Psi_{v+1}$. For $\hat D$ only the bulk-velocity normalization is used, since the displacement normalization fixes $\hat\Delta_{\hat D}=2$ by construction; its error is the half-spread of the two- and three-parameter fits.}
\label{tab:dims}
\end{table*}

\begin{table*}[h]
\centering\footnotesize
\setlength{\tabcolsep}{5pt}
\begin{tabular}{l|l|c|c|c|c|c|c}
\hline
\hline 
quantity & & $v=1/12$ & $v=1/6$ & $v=1/4$ & $v=1/3$ & $v=5/12$ & $v=1/2$\\
\hline
$b_{\phi\Psi_v}$ & FS & 0.990(16) & 0.977(14) & 0.957(12) & 0.933(10) & 0.905(9) & 0.876(7)\\
 & free & 0.949 & 0.907 & 0.873 & 0.844 & 0.820 & 0.798\\
\hline
$b_{\phi\Psi_{v-1}}$ & FS & 0.766(10) & 0.781(8) & 0.799(7) & 0.822(6) & 0.848(6) & 0.876(7)\\
 & free & 0.719 & 0.732 & 0.746 & 0.761 & 0.779 & 0.798\\
\hline
$C_{\Psi_v\Psi_v(2,2v)}$ (GFF $\sqrt2$) & FS & 1.333(74) & 1.339(68) & 1.349(61) & 1.362(53) & 1.376(44) & 1.390(35)\\
\hline
$C_{\Psi_v\bar\Psi_{1-v}\hat D}$ (GFF 1) & FS & 0.820(27) & 0.894(27) & 0.939(33) & 0.959(41) & 0.966(48) & 0.962(54)\\
\hline
$C_{\Psi_v\bar\Psi_{-v}\mathcal N_1}$ & FS & 0.868(41) & 0.854(40) & 0.833(39) & 0.800(38) & 0.746(34) & 0.672(21)\\
\hline
$C_{\Psi_{v-1}\bar\Psi_{1-v}\mathcal N_1}$ & FS & 0.241(81) & 0.301(77) & 0.375(69) & 0.470(55) & 0.586(32) & 0.710(3)\\
\hline
$b_{S\mathcal N_1}$ & FS & 1.023(79) & 0.962(75) & 0.842(65) & 0.645(48) & 0.357(24) & 0.000(3)\\
\hline
\hline 
\end{tabular}
\caption{Bulk-to-defect couplings (at the equator) and defect OPE coefficients of $\mathcal D_v$, extrapolated from $\Norb=10$--$13$, with the
free-theory couplings and the GFF values of the OPE coefficients. At $v=\frac12$ the two coefficients of $\mathcal N_1$ must coincide in magnitude by
$R_y(\pi)$, so that their difference measures the systematic error of the ratio method, and $b_{S\mathcal N_1}$ vanishes by charge
conjugation.}
\label{tab:ope}
\end{table*}

\section{The $v\to0^+$ crossover}
\label{app:smallv}
The finite-size construction is exactly periodic in $v$ for non-integer $v$ but is not continuous at $v\to0$: at $v=0$ the charged flavours
have the $2s+1$ orbitals of the bulk, while for $0<v<1$ each has $2s$, with the wavefunctions $r^{v}$ and $r^{1-v}$ at the poles. At
$v\to0^+$ the Hamiltonian is therefore that of the trivial defect with the $m=s_M$ orbital of $\psi_+$ and the $m=-s_M$ orbital of $\psi_-$ removed, a strong local
perturbation at the poles that decays only as a power of $\Norb$ [Fig.~\ref{fig:cross}]: at $v=10^{-3}$ and $\Norb=10$ the lowest charged
level lies $0.17$ above $\Delta_\phi(\Norb)$ and the lowest neutral level $0.20$ above $\Delta_S(\Norb)$, with fitted decay exponents of
$0.32$--$0.44$ depending on the fit window. The fixed-order one-loop expansion is not uniform in $v$ either; the resummation of Sec.~\ref{app:oneloop} restores the correct $v\to0$ limit of the lowest charged dimension but does not remove the finite-size crossover of the microscopic construction. The strongest crossover effects are observed in the lowest charge-one and neutral sectors. Charge-two composites involving these modes, such as $(2,2v)$, can inherit small-$v$ corrections; the smooth finite-size sequences of the other operators alone do not establish uniform convergence to the transparent-defect limit. A cleaner small-$v$ regime would require compensating the pole perturbation by a local counterterm or
substantially larger systems.

\begin{figure}[t]
\centering
\includegraphics[width=0.5\linewidth]{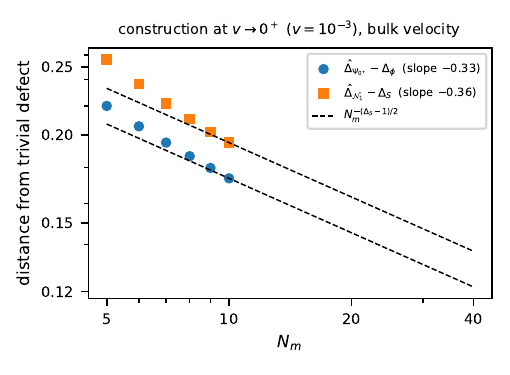}
\caption{Distance of the lowest charged and neutral levels of the $v\to0^+$ construction ($v=10^{-3}$, $\Norb=5$--$10$) from the
corresponding levels of the trivial defect, versus $\Norb$ (bulk velocity). The legend gives the fitted slopes; the dashed lines show the power law $\Norb^{-(\Delta_S-1)/2}$ expected if the leading correction is governed by the bulk singlet restricted to the transparent line.}
\label{fig:cross}
\end{figure}

\end{widetext}

\end{document}